\documentclass[aps, pra, twocolumn, amsmath, amssymb, 
superscriptaddress, 
nofootinbib]{revtex4-1}

\usepackage[utf8]{inputenc}
\usepackage{graphicx}
\usepackage{units}
\usepackage{color}
\usepackage[pdftex, colorlinks=true, linkcolor=myblue, citecolor=myblue,
urlcolor=myblue]{hyperref}
\usepackage{comment}
\usepackage{graphicx}
\usepackage{amsmath, calc}
\usepackage{amsfonts}
\usepackage{amssymb}
\usepackage{bm}
\usepackage{bbm}
\usepackage{color}
\usepackage{array}
\usepackage{units}
\definecolor{myblue}{rgb}{0,0,1}
\newcommand{\bra}[1]{\langle #1|}
\newcommand{\ket}[1]{|#1\rangle}

\newcommand{\expect}[1]{\langle #1 \rangle}

\begin{document}

\title{Topological plasmons in dimerized chains of nanoparticles: robustness against long-range quasistatic interactions and retardation effects}

\author{Charles A. Downing}
\affiliation{Departamento de F\'{i}sica T\'{e}orica de la Materia Condensada and Condensed Matter Physics Center (IFIMAC),
Universidad Aut\'{o}noma de Madrid, E-28049 Madrid, Spain}

\author{Guillaume Weick}
\email{guillaume.weick@ipcms.unistra.fr} 
\affiliation{Universit\'{e} de Strasbourg, CNRS, Institut de Physique et Chimie des Mat\'{e}riaux de Strasbourg,  UMR 7504, F-67000 Strasbourg, France}

%\date{\today}

\begin{abstract}
We present a simple model of collective plasmons in a dimerized chain of spherical metallic nanoparticles, an elementary example of a topologically nontrivial nanoplasmonic system. Taking into account long-range quasistatic dipolar interactions throughout the chain, we provide an exact analytical expression for the full quasistatic bandstructure of the collective plasmons. 
An explicit calculation of the Zak phase proves the robustness of the topological physics of the system against the inclusion of long-range Coulomb interactions, despite the broken chiral symmetry. 
Using an open quantum systems approach, which includes retardation through the plasmon-photon coupling, we go on to analytically evaluate the resulting radiative frequency shifts of the plasmonic spectrum. 
The bright plasmonic bands experience size-dependent radiative shifts, while the dark bands are essentially unaffected by the light-matter coupling. Notably, the upper transverse-polarized band presents a logarithmic singularity where the quasistatic spectrum intersects the light cone. At wavevectors away from this intersection and for subwavelength nanoparticles, the plasmon-photon coupling only leads to a quantitative  reconstruction of the bandstructure and the topologically-protected states at the edge of the first Brillouin zone are essentially unaffected. 
\end{abstract}

\maketitle

%==================================================
%==================================================
%==================================================
%==================================================
\section{Introduction}
\label{Sec:intro}

The simple system of a chain of regularly-spaced sites coupled by short-range interactions is an insightful model across many areas of physics. Dimerized chains are of only a minor additional complexity, but present significant differences in the physics manifested in the system. For example, Peierls made the profound discovery in the 1930's that the distortion of a one-dimensional atomic chain may be energetically favorable, leading to the celebrated Peierls transition \cite{Peierls1955}. While modeling electrons hopping along the alternating single and double C--C bonds of polyacetylene, Su, Schrieffer and Heeger wrote down a physically rich Hamiltonian which encompasses solitons and charge fractionalization \cite{Su1979, Su1980, Heeger1988}. More recently, Kitaev proposed an influential toy model which supports unpaired Majorana modes at the end of a quantum wire~\cite{Kitaev2001}.
 
The field of plasmonics \cite{Maier2007,Stockman2011} aims to control and manipulate plasmons, the collective oscillations of the electron gas, in order to facilitate technological applications at the nanoscale, including, e.g.\ sensing and waveguiding~\cite{Halas2007}. The localized surface plasmon (LSP) is a particularly promising quasiparticle to exploit due to advances in the controlled synthesis of metallic nanoparticles, which host such a collective excitation~\cite{Bertsch, Kreibig}.

Various arrays of nanoparticles have been investigated \cite{Meinzer2014, Wang2018}, where Coulomb interactions between the nanoparticles give rise to collective plasmons spread out over the whole metamaterial. The simplest array hosting collective plasmons is the linear chain of metal nanoparticles, which is the prototypical plasmonic waveguide~\cite{Quinten1998}. The regular plasmonic chain has been extensively studied both theoretically 
\cite{Brongersma2000, Maier2003b, Park2004, Weber2004, Citrin2004, Simovski2005, Citrin2006, Koenderink2006, Markel2007, Fung2007, Lee2012, Compaijen2013, Pino2014, Compaijen2015, Petrov2015, Brandstetter2016, Downing2018, Compaijen2018}
and experimentally~\cite{Krenn1999, Maier2002, Maier2003a, Koendrick2007, Crozier2007, Apuzzo2013, Barrow2014, Gur2017}. Crucially, it was noticed theoretically that retardation in the dipole-dipole interaction amongst the LSPs can have a dramatic influence on the plasmonic bandstructure~\cite{Weber2004, Citrin2004, Simovski2005, Citrin2006, Koenderink2006, Markel2007, Fung2007, Compaijen2015, Petrov2015, Downing2018, Compaijen2018}. In particular, it was demonstrated that plasmonic modes polarized perpendicular to the chain, and with wavenumbers corresponding to the intersection of the quasistatic plasmonic dispersion and the light cone, experience a significant radiative redshift. 

The inevitable marriage of dimerized chains and nanoplasmonics was studied theoretically in recent years, primarily at the level of nearest-neighbor interactions and neglecting retardation effects \cite{Ling2015, DowningBiPartite2017, Gomez2017}. Within the two above-mentioned approximations, it was shown that a bipartite chain of plasmonic nanoparticles should host midgap topological edge states which are robust against disorder. Zigzag chains of plasmonic particles offer an alternative to the dimerized geometry and were studied both theoretically \cite{Poddubny2014} and experimentally~\cite{Sinev2015}.

Nonplasmonic one-dimensional photonic systems such as photonic lattices \cite{Kanshu2012, Schomerus2013}, dielectric resonators \cite{Poli2015, Slobo2015}, silicon waveguides \cite{Blanco2016}, polariton cavities \cite{Solny2016, Solny2016b}, metallic nanowires \cite{liu16_preprint}, and microring resonators \cite{Zhao2018, Parto2018} were also demonstrated as being the hosts of nontrivial topological states of light. Higher-dimensional topological artificial systems are also at present intensively studied, leading to many possible exciting applications in photonics (for recent reviews, see references \cite{Lu2014, Ozawa2018}).

Recently, Pocock \textit{et al.}\ \cite{Pocock2018} legitimately challenged the findings of references \cite{Ling2015, DowningBiPartite2017, Gomez2017} against the long-ranged nature of the quasistatic dipole-dipole interaction, as well as retardation effects and radiation damping stemming from the finite velocity of light. Based on numerical solutions to Maxwell's equations, it was argued that a nontrivial Zak phase \cite{Berry1984, Zak1989, Asboth2016}, which dictates the existence of topologically-protected edge states via the bulk-edge correspondence \cite{Hasan2010, Delplace2011, Rhim2017}, persists even in the presence of such long-ranged, retarded interactions. 

In this work, we extend the theory of the dimerized plasmonic chain with a transparent model, built within an open quantum system approach \cite{Brandstetter2015, Brandstetter2016}. In our model, we account for long-range quasistatic interactions exactly, which enables us to obtain an analytical expression for the plasmonic bandstructure. By analyzing the properties of the Bloch Hamiltonian, we show that despite the broken chiral (or sublattice) symmetry, time-reversal and inversion symmetries are preserved, guaranteeing a quantized Zak phase \cite{vanMiert2017} which we calculate exactly. 

We explicitly include retardation through the light-matter minimal coupling Hamiltonian, which we account for perturbatively. Our analytical treatment is advantageous as compared to a numerical classical approach~\cite{Pocock2018}, which requires the introduction of a radiative correction to the polarizability via a term which can lead to acausal behavior \cite{Weber2004, Jackson1998}. Our modus operandi here is underpinned by the fluctuation-dissipation theorem \cite{Kubo1966}: the photonic environment leads to both a finite radiative lifetime and a shift in the plasmonic energy levels \cite{Downing2017}, in direct analogy with the famous Lamb shift of atomic physics \cite{Lamb1947, Bethe1947, Milonni1994}. The radiative frequency shifts calculated here are a manifestation of the cooperative Lamb shift, well known from many-atom systems \cite{Friedberg1973, Scully2009}, and give rise to insignificant modifications to the plasmonic bandstructure for small-enough nanoparticles, thus preserving the existence of the topologically-protected edge states.  

The paper is organized as follows: We present our model of a dimerized chain of plasmonic nanoparticles coupled to vacuum photonic modes in section \ref{Sec:model}, and we derive a simple expression for the exact quasistatic plasmonic bandstructure in section \ref{sec:qsa}. In section \ref{Sec:topological}, we comment on the robustness of the topologically nontrivial physics of the system. We unveil analytical expressions for the radiative shifts of the collective plasmonic bandstructure in section \ref{Sec:shifts}. Finally, we draw some conclusions in section \ref{sec:conc}. 
Details of our calculations, as well as complementary analytical results on the radiative damping decay rates of the system, are presented in the appendices.

%==================================================
%==================================================
%==================================================
%==================================================
\section{Model}
\label{Sec:model}

\begin{figure}[tb]
 \includegraphics[width=\linewidth]{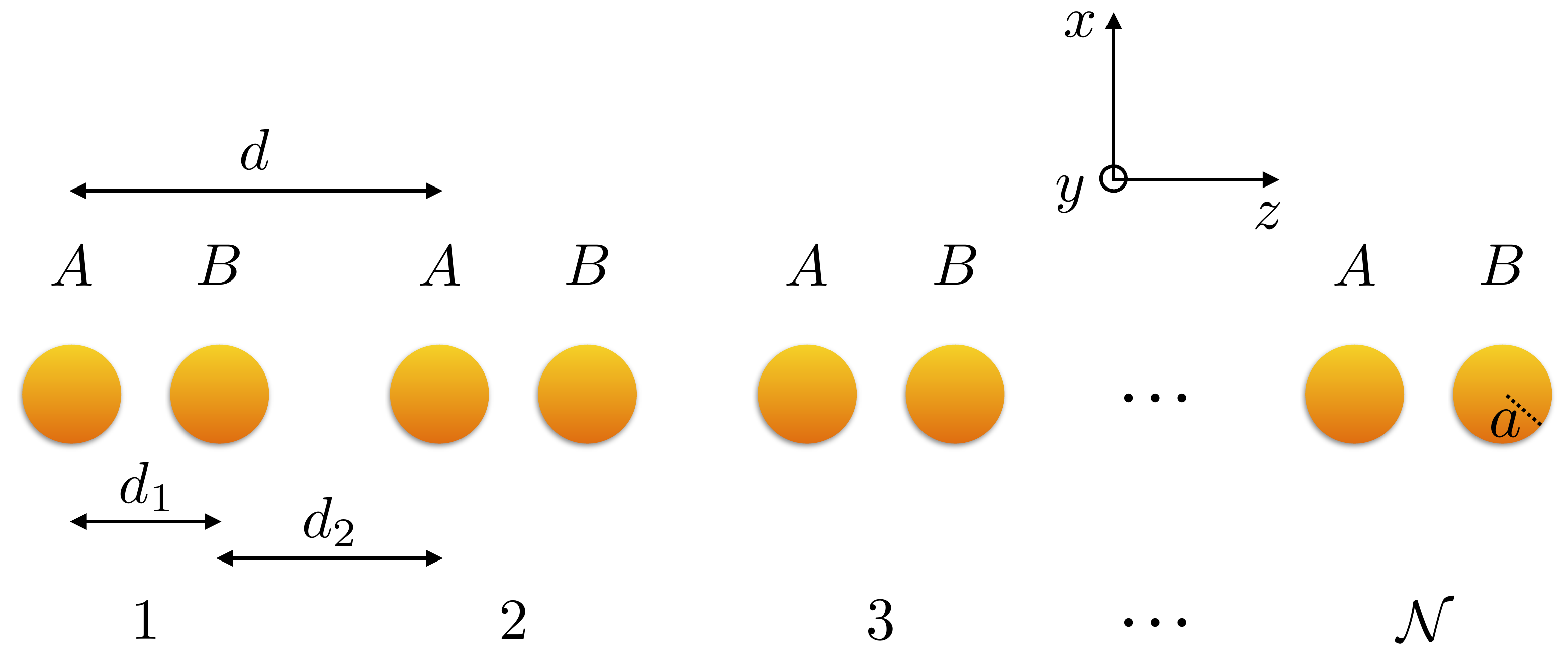}
 \caption{Sketch of a bipartite chain of period $d$, comprised of $\mathcal{N}$ dimers of spherical metallic nanoparticles. The nanoparticles, each of radius $a$, belong to either the $A$ or $B$ sublattices and are separated by the alternating center-to-center distances $d_1$ and $d_2$. }
 \label{fig:sketch}
\end{figure}

We consider a dimerized one-dimensional array of spherical metallic nanoparticles, each of radius $a$, characterized by two inequivalent sublattices which we denote $A$ and $B$ (see figure \ref{fig:sketch}). The formed bipartite chain is characterized by two interparticle separations, $d_1$ and $d_2$, which alternate along the array and hence give rise to the periodicity $d = d_1 + d_2$. 
Every nanoparticle harbors three degenerate LSPs, each polarized along one of the two transverse directions $x$ and $y$, or along the longitudinal direction $z$. The LSP corresponds to an oscillation of the electronic center of mass at the Mie frequency $\omega_0$~\cite{Bertsch, Kreibig}.  For nanoparticles in vacuum, and neglecting the screening of the valence electrons by the core electrons, the Mie frequency corresponds to $\omega_{\mathrm{p}} / \sqrt{3}$, where $\omega_\mathrm{p}=\sqrt{4\pi n_\mathrm{e}e^2/m_\mathrm{e}}$ is the plasma frequency. Here, $n_\mathrm{e}$ is the electronic density of the considered metal, $-e<0$ the electronic charge, and $m_\mathrm{e}$ its mass.\footnote{Throughout the paper, we use cgs units.} 

Crucially, the LSPs in the chain couple to form collective plasmons spread out over the whole chain due to Coulomb interactions, which are essentially dipolar for the interparticle separations $d_{1, 2} \gtrsim 3 a$ we consider here~\cite{Park2004}. 
Retardation effects, accounted for perturbatively in our model by the coupling of the plasmonic modes to the photonic environment, lead to a reconstruction of the resulting collective plasmonic bandstructure. 

Within the Coulomb gauge  \cite{cohen, craig}, the fully-retarded Hamiltonian of the plasmonic chain coupled to vacuum electromagnetic modes is
\begin{equation}
\label{eq:Ham}
H = H_{\mathrm{pl}} + H_\mathrm{ph} + H_{\mathrm{pl}\textrm{-}\mathrm{ph}},
\end{equation}
where $H_{\mathrm{pl}}$ and $H_\mathrm{ph}$ respectively describe the electronic and photonic subsystems coupled through the light-matter interaction $H_{\mathrm{pl}\textrm{-}\mathrm{ph}}$.\footnote{In contrast to reference \cite{DowningBiPartite2017}, in this work we focus on radiative effects and neglect Landau damping~\cite{Bertsch, Kreibig, Kawabata1966, Weick2005}, and hence the associated shift it induces in the plasmonic resonance frequency~\cite{Weick2006}. This is justified as long as the nanoparticles are not too tiny (i.e.\ those with a radius $a \lesssim \unit[5]{nm}$), rendering quantum-size effects irrelevant.}

The plasmonic Hamiltonian, describing the LSPs coupled via the long-ranged quasistatic dipolar interaction, reads
\begin{equation}
\label{eq:Ham_chain}
  H_{\mathrm{pl}} =   H_{\mathrm{pl}}^0+H_{\mathrm{pl}}^\mathrm{int},
\end{equation}
where the noninteracting part
\begin{equation}
\label{eq:Ham_chain_0}
H_{\mathrm{pl}}^0 = \hbar \omega_0  \sum_{\sigma=x,y,z}\sum_{n=1}^\mathcal{N}\left( {a_{n}^{\sigma}}^{\dagger} a_{n}^{\sigma} + {b_{n}^{\sigma}}^{\dagger} b_{n}^{\sigma} \right)
\end{equation}
 corresponds to the Hamiltonian for uncoupled harmonic LSPs.
Here, the index $n \in [1, \mathcal{N}]$ denotes the dimer number in the chain of $2 \mathcal{N}$ nanoparticles\footnote{Without loss of generality, we consider chains with an even number of nanoparticles.}
and $\sigma$ labels the two transverse ($x, y$) and the single longitudinal ($z$) polarizations of the plasmonic modes 
(see figure \ref{fig:sketch}). In equation \eqref{eq:Ham_chain_0}, the bosonic operators $a_{n}^{\sigma}$ (${a_{n}^{\sigma}}^{\dagger}$) and $b_{n}^{\sigma}$ (${b_{n}^{\sigma}}^{\dagger}$) annihilate (create) an LSP with polarization $\sigma$ in dimer $n$ in a nanoparticle belonging to the $A$ and $B$ sublattices, respectively. 

In equation \eqref{eq:Ham_chain}, the long-range quasistatic dipolar interaction among the LSPs is encapsulated in the term $H_{\mathrm{pl}}^\mathrm{int}$, which can be decomposed into two contributions: 
\begin{equation}
\label{eq:H_pl_decompo}
H_{\mathrm{pl}}^\mathrm{int}=H_{\mathrm{pl}}^{A\leftrightarrow A, B\leftrightarrow B}+H_{\mathrm{pl}}^{A\leftrightarrow B}.
\end{equation}
The first of these contributions corresponds to the coupling between pairs of nanoparticles belonging to the same sublattices ($A \leftrightarrow A$ and $B \leftrightarrow B$) and reads
\begin{align}
\label{eq:H_AA}
H_{\mathrm{pl}}^{A\leftrightarrow A, B\leftrightarrow B}=&\;\hbar\Omega\sum_{\sigma=x,y,z}\sum_{n=1}^\mathcal{N}
\sum_{m=n+1}^\mathcal{N}\frac{\eta_\sigma}{(m-n)^3}
\nonumber\\
&\times
\left[
\left({a_n^\sigma}^\dagger+a_n^\sigma\right)\left({a_m^\sigma}^\dagger+a_m^\sigma\right)
\right.
\nonumber\\
&+\left.
\left({b_n^\sigma}^\dagger+b_n^\sigma\right)\left({b_m^\sigma}^\dagger+b_m^\sigma\right)
\right].
\end{align}
The second contribution $H_{\mathrm{pl}}^{A\leftrightarrow B}$ to equation \eqref{eq:H_pl_decompo} is due to the coupling between pairs of particles belonging to the two inequivalent sublattices ($A\leftrightarrow B$). It can be written as
\begin{align}
\label{eq:H_AB}
H_{\mathrm{pl}}^{A\leftrightarrow B}=&\;\hbar\Omega\sum_{\sigma=x,y,z}\sum_{n=1}^\mathcal{N}
\Bigg\{
\sum_{m=n}^\mathcal{N}\frac{\eta_\sigma}{(m-n+d_1/d)^3}
\nonumber\\
&\times\left({a_n^\sigma}^\dagger+a_n^\sigma\right)\left({b_m^\sigma}^\dagger+b_m^\sigma\right)
\nonumber\\
&+\sum_{m=n+1}^\mathcal{N}\frac{\eta_\sigma}{(m-n-1+d_2/d)^3}
\nonumber\\
&\times\left({b_n^\sigma}^\dagger+b_n^\sigma\right)\left({a_m^\sigma}^\dagger+a_m^\sigma\right)
\Bigg\}.
\end{align}
In equations \eqref{eq:H_AA} and \eqref{eq:H_AB}, the coupling constant $\Omega$, defined by
\begin{equation}
\label{eq:Coupling}
\Omega = \frac{\omega_0}{2} \left( \frac{a}{d} \right)^3,
\end{equation}
is much smaller than $\omega_0$ for all realistic values of the ratio $d/a$,\footnote{Note that touching nanoparticles would correspond to $d/a=4$, i.e.\ $\Omega/\omega_0=1/128$.} while
the polarization-dependent factor $\eta_{x,y} = 1$ ($\eta_{z} = -2$) for the transverse (longitudinal) modes arises from the anisotropy of the dipole-dipole interaction.

We note that the nearest-neighbor approximation adopted in reference \cite{DowningBiPartite2017} 
amounts to two simplifications in the interacting Hamiltonian \eqref{eq:H_pl_decompo}. Firstly, one
disregards the intrasublattice Hamiltonian \eqref{eq:H_AA}. Secondly,  one retains in the summation over the dimer index $m$ only the contributions $m=n$ and $m=n+1$ in the first and second terms on the right-hand side of the intersublattice Hamiltonian \eqref{eq:H_AB}, respectively (cf.\ equation (2) in reference \cite{DowningBiPartite2017}). 

In equation \eqref{eq:Ham}, the photonic environment is described by the Hamiltonian
\begin{equation}
\label{eq:H_ph}
 H_{\mathrm{ph}} = \sum_{\mathbf{k}, \hat{\lambda}_{\mathbf{k}}} \hbar \omega_{\mathbf{k}} {c_{\mathbf{k}}^{\hat{\lambda}_{\mathbf{k}}}}^{\dagger} c_{\mathbf{k}}^{\hat{\lambda}_{\mathbf{k}}}, 
\end{equation}
where $c_{\mathbf{k}}^{\hat{\lambda}_{\mathbf{k}}}$ (${c_{\mathbf{k}}^{\hat{\lambda}_{\mathbf{k}}}}^{\dagger}$) annihilates (creates) a photon with wavevector $\mathbf{k}$ and transverse polarization $\hat{\lambda}_{\mathbf{k}}$, so that $\mathbf{k}\cdot\hat{\lambda}_{\mathbf{k}}=0$ (here and in what follows, hats designate unit vectors). The photonic dispersion is $\omega_{\mathbf{k}} = c |\mathbf{k}|$, where $c$ is the speed of light in vacuum. 

In the long-wavelength approximation $|\mathbf{k}| a \ll1$, the plasmon-photon coupling Hamiltonian in equation \eqref{eq:Ham} is
\begin{align}
\label{eq:HamCoupling}
H_{\mathrm{pl}\textrm{-}\mathrm{ph}} =&\;  \sum_{n=1}^{\mathcal{N}} \sum_{s=A, B} \bigg\{ \frac{e}{m_{\mathrm{e}}c} \mathbf{\Pi}_{n, s} \cdot \mathbf{A} (\mathbf{d}_{n, s}) 
 \nonumber\\
&+ \frac{N_{\mathrm{e}} e^2}{2 m_{\mathrm{e}}c^2} \mathbf{A}^2 (\mathbf{d}_{n, s}) \bigg\},
\end{align}
where $\mathbf{d}_{n, A} = d (n-1) \hat{z}$ and $\mathbf{d}_{n, B} = \mathbf{d}_{n, A} + d_1 \hat{z}$ correspond, respectively, to the location of the center of the nanoparticle belonging to the $A$ and $B$ sublattices in the dimer $n$ (see figure~\ref{fig:sketch}). The momenta associated with the LSPs on the $A$ and $B$ sublattices are 
\begin{subequations}
\begin{align}
\mathbf{\Pi}_{n,A} = \mathrm{i}\sqrt{\frac{N_\mathrm{e}m_\mathrm{e}\hbar\omega_0}{2}}\sum_{\sigma=x,y,z} \hat\sigma~({a_n^\sigma}^\dagger-a_n^\sigma),
\\
\mathbf{\Pi}_{n,B} = \mathrm{i}\sqrt{\frac{N_\mathrm{e}m_\mathrm{e}\hbar\omega_0}{2}}\sum_{\sigma=x,y,z} \hat\sigma~({b_n^\sigma}^\dagger-b_n^\sigma).
\end{align}
\end{subequations}
The vector potential appearing in equation \eqref{eq:HamCoupling} is given by
\begin{equation}
\label{eq:A}
\mathbf{A}(\mathbf{d}_{n, s})=\sum_{\mathbf{k}, \hat\lambda_{\mathbf{k}}}
\hat\lambda_{\mathbf{k}}\sqrt{\frac{2\pi\hbar c^2}{\mathcal{V}\omega_\mathbf{k}}}
\left(
c_\mathbf{k}^{\hat\lambda_{\mathbf{k}}}\mathrm{e}^{\mathrm{i}\mathbf{k}\cdot \mathbf{d}_{n, s}}
+{c_\mathbf{k}^{\hat\lambda_{\mathbf{k}}}}^\dagger\mathrm{e}^{-\mathrm{i}\mathbf{k}\cdot \mathbf{d}_{n, s}}
\right),
\end{equation}
where we have quantized the electromagnetic modes in a box of volume $\mathcal{V}$. Notably, the Hamiltonian \eqref{eq:HamCoupling} fully takes into account retardation effects in the dipole-dipole interaction between the LSPs \cite{craig, Downing2018, Lamowski2018}.\footnote{Note that, since we consider interparticle separation distances much smaller than the wavelength associated with the LSP resonances, we neglect Umklapp processes in equations \eqref{eq:H_ph} and \eqref{eq:HamCoupling}.}

%==================================================
%==================================================
%==================================================
%==================================================
\section{Quasistatic plasmonic bandstructure}
\label{sec:qsa}

We begin our analysis of the model presented in section \ref{Sec:model} by considering the purely plasmonic Hamiltonian \eqref{eq:Ham_chain} and deriving its associated quasistatic spectrum, thus gaining an understanding of the bandstructure without the influence of retardation effects. 

Since we are interested in long chains where $\mathcal{N} \gg 1$,\footnote{Such a limit has been shown to be a good approximation for chains comprising $20$ or more nanoparticles~\cite{Weber2004, Brandstetter2016, Downing2018}.}
it is expedient to employ periodic boundary conditions. We move into wavevector space via the pair of Fourier transforms 
\begin{subequations}
\label{eq:basis}
\begin{align}
 a_{n}^{\sigma} &= \frac{1}{\sqrt{\mathcal{N}}} \sum_{q} \mathrm{e}^{\mathrm{i} n q d} a_{q}^{\sigma}, \\
 b_{n}^{\sigma} &= \frac{1}{\sqrt{\mathcal{N}}} \sum_{q} \mathrm{e}^{\mathrm{i} n q d} b_{q}^{\sigma}, 
 \end{align}
\end{subequations}
where the plasmonic wavevector $q =  2 \pi p /\mathcal{N} d$, with the integer $p\in[-\mathcal{N}/2, +\mathcal{N}/2]$.
Hence the plasmonic Hamiltonian \eqref{eq:Ham_chain} becomes
\begin{align}
\label{eq:Ham_transformed}
  H_{\mathrm{pl}} =&\; \sum_{\sigma q} \bigg\{ \left( \hbar\omega_0 + \hbar\Omega f_{q}^{\sigma} \right) \left( {a_{q}^{\sigma}}^{\dagger} a_{q}^{\sigma} + {b_{q}^{\sigma}}^{\dagger} b_{q}^{\sigma} \right) \nonumber\\
  &+ \frac{\hbar \Omega}{2}  f_q^{\sigma} \left( {a_{q}^{\sigma}}^{\dagger} {a_{-q}^{\sigma}}^{\dagger} +  a_{-q}^{\sigma} a_q^{\sigma} + {b_{q}^{\sigma}}^{\dagger} {b_{-q}^{\sigma}}^{\dagger} +  b_{-q}^{\sigma}b_q^{\sigma}  \right)   \nonumber\\
&+ \hbar \Omega  \left[ g_{q}^{\sigma} {a_{q}^{\sigma}}^{\dagger} \left( b_q^{\sigma} + {b_{-q}^{\sigma}}^{\dagger} \right)
+{g_{q}^{\sigma}}^\ast {a_{q}^{\sigma}} \left( {b_{q}^{\sigma}}^\dagger + {b_{-q}^{\sigma}}\right)\right]    \bigg\}, 
 \end{align}
where we have introduced the lattice sums
\begin{equation}
\label{eq:f_q_def}
f_{q}^{\sigma} = 2\eta_\sigma \sum_{n=1}^\infty \frac{\cos{(nqd)}}{n^3}
\end{equation}
and 
\begin{equation}
\label{eq:g_q_def}
g_{q}^{\sigma} = \eta_\sigma \sum_{n=0}^\infty \left[ \frac{\mathrm{e}^{\mathrm{i} n q d}}{(n+d_1/d)^3} + \frac{\mathrm{e}^{-\mathrm{i} (n+1) q d}}{(n+d_2/d)^3} \right], 
\end{equation}
which correspond to the intrasublattice ($A\leftrightarrow A$ and $B\leftrightarrow B$)  
and intersublattice ($A\leftrightarrow B$) interactions, respectively. 
The latter quantities can be expressed analytically in terms of the polylogarithm function $\text{Li}_s (z)=\sum_{n=1}^{\infty} z^n n^{-s}$ and the Lerch transcendent $\Phi (z, s, u) = \sum_{n=0}^{\infty} z^n(n+u)^{-s}$ as follows:
\begin{equation}
\label{eq:fsum} 
 f_{q}^{\sigma} =  \eta_{\sigma} \left[ \mathrm{Li}_3 \left( \mathrm{e}^{\mathrm{i} q d} \right) 
 + \mathrm{Li}_3 \left( \mathrm{e}^{-\mathrm{i} q d} \right) \right] 
 \end{equation}
 and 
 \begin{equation}
 \label{eq:gsum}
  g_{q}^{\sigma} =  \eta_{\sigma} \left[  \Phi \left( \mathrm{e}^{\mathrm{i} q d}, 3, d_1/d \right) 
 + \mathrm{e}^{-\mathrm{i} q d}\, \Phi \left( \mathrm{e}^{-\mathrm{i} q d}, 3, d_2/d \right) \right].
 \end{equation}
The behavior of the intra-  and intersublattice functions, respectively equations \eqref{eq:fsum} and \eqref{eq:gsum}, is analyzed in detail in appendix~\ref{app:lattice_sums}. 

Since the plasmonic Hamiltonian \eqref{eq:Ham_transformed} is quadratic, it can be diagonalized exactly with the aid of a bosonic Bogoliubov transformation. We arrive at the form $H_{\mathrm{pl}} = \sum_{\sigma q \tau} H_{q\tau}^\sigma$, where
\begin{equation}
\label{eq:plchain}
H_{q\tau}^\sigma=  \hbar \omega_{q \tau}^{\sigma} {\beta_{q \tau}^{\sigma}}^{\dagger} \beta_{q \tau}^{\sigma},
\end{equation}
with the index $\tau = +$ ($-$) corresponding to the upper (lower) collective plasmonic band. The eigenfrequencies $\omega_{q \tau}^{\sigma}$ of the collective plasmonic modes are 
\begin{equation}
\label{eq:plspectrum}
\omega_{q \tau}^{\sigma} = \omega_0 \sqrt{1 + 2 \frac{\Omega}{\omega_0} \left( f_q^{\sigma} + \tau |g_q^{\sigma}| \right)},
\end{equation}
which provides a fully analytical description of the quasistatic plasmonic bandstructure. 

The bosonic Bogoliubov operators appearing in the Hamiltonian~\eqref{eq:plchain} are given by
\begin{equation}
\label{eq:Bog}
\beta_{q \tau}^{\sigma} = \cosh{\theta_{q\tau}^{\sigma}}\ \alpha_{q\tau}^{\sigma} + \sinh{\theta_{q\tau}^{\sigma}}\ {\alpha_{-q\tau}^{\sigma}}^{\dagger}, 
\end{equation}
with 
\begin{equation}
\label{eq:op}
 \alpha_{q\tau}^{\sigma} = \frac{1}{\sqrt{2}}  \left( a_q^{\sigma} + \tau \frac{g_q^{\sigma}}{|g_q^{\sigma}|} b_q^{\sigma} \right), 
\end{equation} 
where the coefficients are
\begin{subequations}
\label{eq:BogCoeff}
\begin{equation}
\label{eq06b}
  \cosh{\theta_{q\tau}^{\sigma}} = \frac{\omega_{q\tau}^\sigma+\omega_0}{2\sqrt{\omega_0\omega_{q\tau}^\sigma}},
\end{equation}
\begin{equation}
\label{eq06c}
  \sinh{\theta_{q\tau}^{\sigma}} = \frac{\omega_{q\tau}^\sigma-\omega_0}{2\sqrt{\omega_0\omega_{q\tau}^\sigma}}.
  \end{equation}
\end{subequations}
The Bogoliubov operators $\beta_{q \tau}^{\sigma}$ and ${\beta_{q \tau}^{\sigma}}^\dagger$ act on an eigenstate $|n_{q \tau}^\sigma\rangle$ of the Hamiltonian \eqref{eq:plchain}, which represents a collective plasmon with polarization $\sigma$, wavevector $q$, band index $\tau$, and eigenenergy $\hbar\omega_{q \tau}^\sigma$, as follows: $\beta_{q \tau}^{\sigma}|n_{q \tau}^\sigma\rangle = \sqrt{n_{q \tau}^\sigma}|n_{q \tau}^\sigma-1\rangle$ and ${\beta_{q \tau}^{\sigma}}^\dagger|n_{q \tau}^\sigma\rangle = \sqrt{n_{q \tau}^\sigma+1}|n_{q \tau}^\sigma+1\rangle$, respectively. Here $n_{q \tau}^\sigma$ is a non-negative integer. 

\begin{figure*}[tb]
 \includegraphics[width=1.0\linewidth]{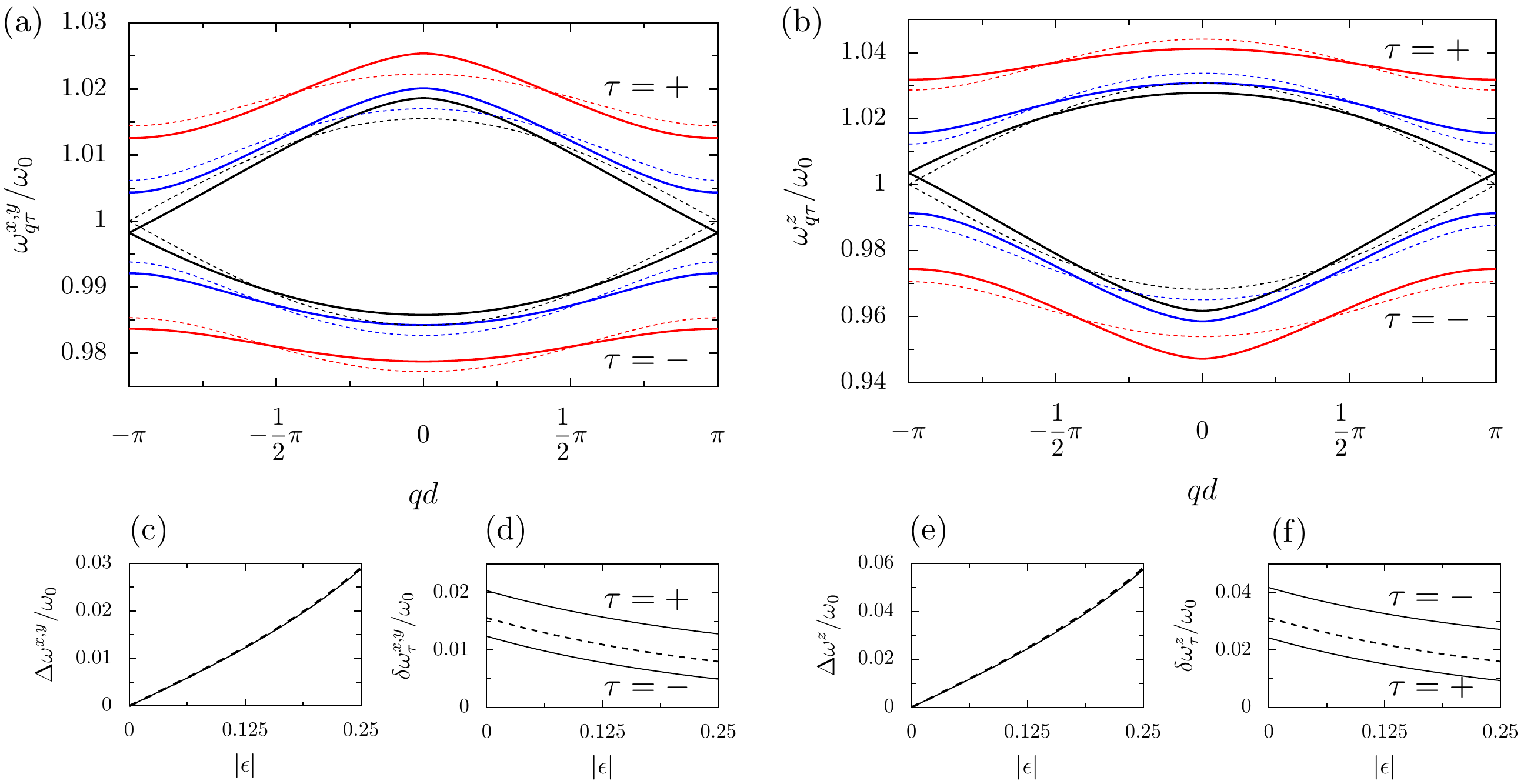}
 \caption{(a),(b) Collective plasmonic bandstructure in the quasistatic limit, in units of the bare LSP resonance frequency $\omega_0$, shown in the first Brillouin zone as a function of the reduced plasmonic wavevector $q d$ and for increasing (absolute) values of the dimerization parameter $|\epsilon|$ (cf.\ equation \eqref{eq:epsilon}): $|\epsilon|=0$ (undimerized case, black lines), $|\epsilon|=1/8$ (blue lines), and $|\epsilon|=1/4$ (red lines). 
 Both the (a) transverse and (b) longitudinal polarizations are displayed. Solid  lines: full quasistatic dispersion relation $\omega_{q \tau}^{\sigma}$, see equation \eqref{eq:plspectrum}. Dashed lines: quasistatic dispersion relation in the nearest-neighbor approximation $\omega_{\mathrm{nn},\,q \tau}^{\sigma}$, see equation \eqref{eq:nearest}. 
(c),(e) Bandgap as a function of the dimerization parameter $|\epsilon|$ for the transverse (panel (c)) and longitudinal modes (panel (e)). 
(d),(f) Bandwidth of the high- ($\tau=+$) and low-energy ($\tau=-$) branches as a function of $|\epsilon|$ for the transverse (panel (d)) and longitudinal modes (panel (f)). 
In panels (c)--(f), the solid (dashed) lines correspond to the full-range quasistatic (nearest-neighbor) results.
 In the figure, the periodicity of the lattice is $d = 8 a$.}
 \label{fig:omega}
\end{figure*}

The quasistatic plasmonic bandstructure of the system, unperturbed by the photonic environment, is analytically encapsulated in equation \eqref{eq:plspectrum}. It is plotted in figure \ref{fig:omega} (see solid lines) for both the transverse (panel (a)) and longitudinal polarizations (panel (b)) for a chain with periodicity $d=8a$ and for different degrees of dimerization, characterized by the parameter $\epsilon$ defined through the relations $d_1=(1+\epsilon)d/2$ and $d_2=(1-\epsilon)d/2$, or equivalently\footnote{For interparticle distances $d_{1,2}\geqslant3a$, the interactions amongst the LSPs are essentially dipolar \cite{Park2004}, which restrict the dimerization parameter in our model to $|\epsilon|\leqslant 1-6a/d$. Note that $\epsilon=0$ corresponds to the regular (undimerized) chain.}
\begin{equation}
\label{eq:epsilon}
\epsilon=\frac{d_1-d_2}{d}.
\end{equation}
In the figure, we also plot as dashed lines the plasmonic dispersion considering only the dipolar interaction between nearest neighbors (nn) in the chain, explicitly (see equation (15) in reference \cite{DowningBiPartite2017})
\begin{equation}
\label{eq:nearest}
\omega_{\mathrm{nn},q \tau}^{\sigma} = \omega_0 \sqrt{1 + 2 \tau \frac{\Omega}{\omega_0} |g_{\mathrm{nn}, q}^{\sigma}| }. 
\end{equation}
Here the quantity which plays the role of the intersublattice function \eqref{eq:gsum}, restricted to nearest neighbors only, is
 \begin{equation}
\label{eq:nearest2}
g_{\mathrm{nn},q}^{\sigma} = \eta_{\sigma} d^3 \left(\frac{1}{d_1^3} + \frac{\mathrm{e}^{-\mathrm{i} q d}}{d_2^3}  \right).
\end{equation}

The weak coupling constant~\eqref{eq:Coupling} implies that the nearest-neighbor spectrum~\eqref{eq:nearest} is essentially
$\omega_{\mathrm{nn},q \tau}^{\sigma} \simeq \omega_0 + \tau \Omega |g_{\mathrm{nn}, q}^{\sigma}|$. Hence $\omega_{\mathrm{nn},\,q \tau}^{\sigma}$ possesses a band reversal symmetry around the resonance frequency $\omega_0$ (or, in the language of solid state physics, a particle-hole symmetry). This symmetry is broken when long-range interactions between LSPs belonging to the same sublattice are included in the model, due to the appearance of the intrasublattice function \eqref{eq:fsum} in equation \eqref{eq:plspectrum}, which also breaks the chiral symmetry of the system \cite{Asboth2016}, such that $\omega_{q \tau}^{\sigma} \simeq \omega_0 + \Omega (f_q^{\sigma} + \tau |g_q^{\sigma}|)$. This raises the question of the impact on the protection of the topological properties of the system. We discuss this important issue in section \ref{Sec:topological}.

As can be seen from the solid lines in panels (c) and (e) of figure \ref{fig:omega} (see also panels (a) and (b)), the bandgap $\Delta\omega^\sigma=\omega^\sigma_{\pi/d,+}-\omega^\sigma_{\pi/d,-}\simeq\omega_0(a/d)^3|g_{\pi/d}^\sigma|$ at the edge of the Brillouin zone increases as a function of 
the dimerization parameter $|\epsilon|$. Such a bandgap is 
almost unaffected by the long-range dipolar interactions, as it very well matches the weak coupling, nearest-neighbor bandgap
\begin{equation}
\Delta\omega^\sigma_\mathrm{nn}=\omega_0|\eta_\sigma| a^3\left|\frac{1}{d_1^3}-\frac{1}{d_2^3}\right|
\end{equation}
represented by a dashed line in figures \ref{fig:omega}(c) and \ref{fig:omega}(e).

It is further apparent from figure \ref{fig:omega} (panels (d) and (f)) that the bandwidth of the high- ($\tau=+$) and low-energy ($\tau=-$) bands $\delta\omega_\tau^\sigma=|\omega_{0,\tau}^\sigma-\omega_{\pi/d,\tau}^\sigma|$, which decreases for increasing $|\epsilon|$, is noticably affected by the long-range interactions. Indeed, while in the weak-coupling, nearest-neighbor approximation, such a bandwidth is $\tau$-independent and given by 
\begin{equation}
\delta\omega_\mathrm{nn}^\sigma=\omega_0|\eta_\sigma|\left(\frac{a}{\mathrm{max}\{d_1,d_2\}}\right)^3, 
\end{equation}
represented by dashed lines in figures \ref{fig:omega}(d) and \ref{fig:omega}(f),
the long-range interactions induce an asymmetry in the bandwith of the high- and low-energy branches (solid lines). 

At the edge of the first Brillouin zone, we find that the collective plasmonic bandstructure~\eqref{eq:plspectrum} maps onto the form of a quasirelativistic spectrum, which indicates the presence of a nontrivial, topologically-protected edge state within the gap~\cite{DowningBiPartite2017}. Performing a Taylor expansion at $qd = \pi + kd$ yields to leading order in $kd \ll 1$
\begin{equation}
\label{eq:edgeBZspectrum}
\omega_{k \tau}^{\sigma} \simeq \omega_{0}^{\sigma} + \tau  \Omega \sqrt{A^{\sigma} + B^{\sigma} ( k d ) ^2},
\end{equation}
where $A^{\sigma}$ and $B^{\sigma}$ are (polarization-dependent) constants, given explicitly in equations \eqref{eq:coeffA} and \eqref{eq:coeffB} in appendix \ref{app:mapping}, and the renormalized frequency
$\omega_0^{\sigma} = \omega_0 - 3\zeta (3) \eta_{\sigma}\Omega/2$,
where $\zeta (3) \simeq 1.20$ is Ap\'{e}ry's constant. Notably, $\omega_{0}^{\sigma}$, which reduces in the nearest-neighbor approximation to $\omega_0$ \cite{DowningBiPartite2017}, is polarization dependent due to the long-range interactions. Up to the constant energy shift $\hbar \omega_{0}^{\sigma}$, the dispersion~\eqref{eq:edgeBZspectrum} recalls the spectrum of one-dimensional Dirac particles with mass $m$ and momentum $p$, $E_{\mathrm{D}} = \pm \sqrt{ (m c^2)^2 + (pc)^2}$. Previously, this mapping was known to hold only in the nearest-neighbor approximation (see equation (16) in reference \cite{DowningBiPartite2017}).

%==================================================
%==================================================
%==================================================
%==================================================
\section{Robustness of the topological properties of the collective plasmons within the quasistatic limit}
\label{Sec:topological}

Here we investigate the robustness of the nontrivial topological nature of the dimerized plasmonic chain against long-range quasistatic interactions. To do so, we analyze the symmetries of the plasmonic Bloch Hamiltonian and
calculate exactly the Berry phase \cite{Berry1984} specialized to one-dimensional systems, the so-called Zak phase~\cite{Zak1989, Asboth2016}.

We start by writing the plasmonic Hamiltonian \eqref{eq:Ham_transformed} in Bogoliubov--de Gennes form as 
$H_\mathrm{pl}=\frac 12 \sum_{\sigma q}\hat{\Psi_q^\sigma}^\dagger \mathcal{H}_q^\sigma \hat\Psi_q^\sigma$, 
with the Bloch Hamiltonian
\begin{equation}
\label{eq:BlochHam}
\mathcal{H}_q^\sigma = \hbar
\begin{pmatrix}
\omega_0+\Omega f_q^\sigma & \Omega g_q^\sigma & \Omega f_q^\sigma & \Omega g_q^\sigma \\[.15truecm]
\Omega {g_q^\sigma}^* & \omega_0+\Omega f_q^\sigma & \Omega {g_q^\sigma}^* & \Omega f_q^\sigma\\[.15truecm]
\Omega f_q^\sigma & \Omega g_q^\sigma & \omega_0+\Omega f_q^\sigma & \Omega g_q^\sigma\\[.15truecm]
\Omega {g_q^\sigma}^* & \Omega f_q^\sigma & \Omega {g_q^\sigma}^* & \omega_0+\Omega f_q^\sigma
\end{pmatrix},
\end{equation}
and where $\hat\Psi_q^\sigma = ( a_q^{\sigma}, b_q^{\sigma}, a_{-q}^{\sigma \dagger}, b_{-q}^{\sigma \dagger})$. The 
$2\pi/d$-periodic part of the Bloch wavefunction, denoted $\ket{\psi_{q\tau}^{\sigma}}$, satisfies the eigenproblem $\mathcal{H}_q^\sigma \mathcal{J}  \ket{\psi_{q\tau}^{\sigma}} = \hbar \omega_{q \tau}^{\sigma} \ket{\psi_{q\tau}^{\sigma}}$, where the plasmonic dispersion $\omega_{q\tau}^\sigma$ is given by equation \eqref{eq:plspectrum}. Here $\mathcal{J} = \sigma_z \otimes \mathbbm{1}_2$, in terms of the third Pauli matrix $\sigma_z$ and the $2 \times 2$ identity matrix $\mathbbm{1}_2$. This complication is due to the requirement of obeying bosonic statistics \cite{Tsallis1978, Colpa1978}. Explicitly, in terms of the phase $\phi_q^\sigma$ defined via
\begin{equation}
\label{eq:phi}
\mathrm{e}^{\mathrm{i}\phi_q^\sigma} = \frac{g_q^\sigma}{|g_q^\sigma|},
\end{equation}
we obtain
\begin{equation}
\label{eq:states}
 \ket{\psi_{q\tau}^{\sigma}} = \frac{1}{\sqrt{2}}
 \begin{pmatrix}
 \cosh{\theta_{q\tau}^{\sigma}} \\[.15truecm]
 \tau\, \mathrm{e}^{\mathrm{i}\phi_q^\sigma} \cosh{\theta_{q\tau}^{\sigma}} \\[.15truecm] 
 \sinh{\theta_{q\tau}^{\sigma}} \\[.15truecm] 
 \tau\, \mathrm{e}^{\mathrm{i}\phi_q^\sigma} \sinh{\theta_{q\tau}^{\sigma}}
 \end{pmatrix},
\end{equation}
 which may be found via a formal identification with the Bogoliubov coefficients \eqref{eq:Bog}. Notably, $\ket{\psi_{q\tau}^{\sigma}}$  is normalized with respect to the metric $\mathcal{J}$, $\expect{\psi_{q\tau}^{\sigma}|\psi_{q\tau}^{\sigma}}_{\mathcal{J}} = \expect{\psi_{q\tau}^{\sigma}| \mathcal{J} |\psi_{q\tau}^{\sigma}} = 1$. We should also mention that the mathematical structure of the wavefunction~\eqref{eq:states} is reminiscent of the same quantity found in the nearest-neighbor approximation (see equation (21) of reference \cite{DowningBiPartite2017}).
 
It was previously discussed in section \ref{sec:qsa} that the inclusion of both the resonance frequency $\omega_0$ and the intrasublattice term \eqref{eq:H_AA} in the full plasmonic Hamiltonian~\eqref{eq:Ham_chain} leads to broken chiral symmetry (see the discussion after equation \eqref{eq:nearest2}). In mathematical terms, we indeed do not have $\mathcal{C}\mathcal{H}_q^\sigma\mathcal{C} = - \mathcal{H}_q^\sigma$ (chiral symmetry) due to $\omega_0\neq0$ and $f_q^\sigma\neq0$, where $\mathcal{C}=\mathbbm{1}_2 \otimes \sigma_z$ is the $4\times4$ generalized third Pauli matrix. 
Therefore, this raises the issue of the implications for the topological protection of the plasmonic edge states. In order to have a quantized Zak phase, it has been recently shown in reference \cite{vanMiert2017} that a 1D insulator is required to respect both time-reversal and inversion symmetries. The Bloch Hamiltonian~\eqref{eq:BlochHam} indeed satisfies both aforementioned criteria: 
\begin{subequations}
\begin{align}
&\mathcal{H}_q^\sigma = {\mathcal{H}_{-q}^\sigma}^{\ast}\qquad \textrm{(time-reversal symmetry)}, \\ 
&\mathcal{I} \mathcal{H}_q^\sigma \mathcal{I} = {\mathcal{H}_{-q}^\sigma}\qquad \textrm{(inversion symmetry)}, 
\end{align}
\end{subequations}
where the $4\times4$ exchange matrix is $\mathcal{I} = \sigma_x \otimes \sigma_x$, in terms of the first Pauli matrix $\sigma_x$. Hence the Zak phase is a meaningful quantity for the present problem, despite the long-range nature of the dipole-dipole interaction, the presence of nonresonant terms in the interacting part of the plasmonic Hamiltonian \eqref{eq:H_pl_decompo}, and the intrinsic bosonic quasiparticle statistics.
 
Formally, the Zak phase is defined as the integral of the Berry connection $\mathcal{A}_{q}^\sigma=\mathrm{i} \bra{\psi_{q\tau}^{\sigma}} \partial_{q} \ket{\psi_{q\tau}^{\sigma}}_{\mathcal{J}}$ over the first Brillouin zone, 
\begin{equation}
\label{eq:zak}
\vartheta_{\mathrm{Z}}^{\sigma} = \int_{\mathrm{1st\, BZ}} \mathrm{d} q \, \mathcal{A}_{q}^\sigma.
\end{equation}
Here, the Berry connection should be calculated while taking into account the aforementioned metric $\mathcal{J}$. Together with equation \eqref{eq:states}, we find 
$\mathcal{A}_{q}^\sigma = \mathrm{i} \bra{\psi_{q\tau}^{\sigma}}\mathcal{J} \partial_{q} \ket{\psi_{q\tau}^{\sigma}}=
-\frac 12 \partial_q\phi_q^\sigma$. 
Carrying out the integral in equation \eqref{eq:zak} with the above result, and taking care with the principal value of the phase $\phi_q^\sigma$, we find the exact result
\begin{equation}
\label{eq:topo}
\vartheta_{\mathrm{Z}}^{\sigma} =
\begin{cases} 
   0, \quad d_1 < d_2, \\
   \pi, \quad d_1 > d_2,
  \end{cases}
\end{equation}
up to modulo $2\pi$, since the Berry connection $\mathcal{A}_{q}^\sigma$ is not gauge invariant. Thus, the dimerized chain has a trivial Zak phase ($\vartheta_{\mathrm{Z}}^{\sigma} = 0$) when $d_1<d_2$ and a nontrivial one ($\vartheta_{\mathrm{Z}}^{\sigma} = \pi$) when $d_1>d_2$, irrespective of the collective mode polarization $\sigma$. There is a topological phase transition at the crossover point of the regularly-spaced chain $d_1 = d_2$, where the bandgap closes (see figure \ref{fig:omega}). Equation \eqref{eq:topo} confirms the presence of plasmonic edge states in the nontrivial phase with $d_1>d_2$, corresponding to topologically-protected localized collective plasmons at the Brillouin zone edge \cite{DowningBiPartite2017}, as directly follows from the bulk-edge correspondence theorem~\cite{Asboth2016}. Such behavior can be understood by a simple physical picture where plasmonic edge states are exponentially pinned at the unpaired nanoparticles located at the two edges of the chain \cite{Downing2017}.

In this section, we have analytically proved the robustness of bosonic edge states against long-range quasistatic interactions and nonresonant couplings. This general result may be important for studies of generalized Su-Schrieffer-Heeger (SSH) models, where previous works have focused on the inclusion of next- and third-nearest neighbor hopping terms only \cite{Li1990, Li1993, Li1995, Li2014, Perez2018}.
Furthermore, there is a recent rise in interest in non-Hermitian SSH models \cite{Zhu2014, Lieu2018, Yin2018, Gong2018, Yao2018, Martinez2018} for which our results may have implications.

We should also mention that it has been shown recently via sophisticated numerical simulations on finite dimerized chains that plasmonic edge states are further robust against both retardation effects and moderate disorder \cite{Pocock2018}. We comment on how retardation affects the topological behavior of the system at the end of section~\ref{Sec:shifts}.

%==================================================
%==================================================
%==================================================
%==================================================
\section{Radiative frequency shifts of the collective plasmonic bandstructure}
\label{Sec:shifts}

We now consider the effect of retardation in the dipole-dipole interaction amongst the nanoparticles in the bipartite chain on the plasmonic bandstructure. Along the lines of reference \cite{Downing2018}, we treat the light-matter coupling Hamiltonian \eqref{eq:HamCoupling} up to second order in perturbation theory, with the aim of obtaining an analytical understanding of the radiative frequency shift induced by the photonic environment. 

In the long-chain limit ($\mathcal{N}\gg1$), our perturbative calculation yields the renormalized collective plasmon dispersion relation (see appendix \ref{app:shifts} for details) 
\begin{equation}
\label{renorm}
\tilde{\omega}_{q \tau}^{\sigma} = \omega_{q \tau}^{\sigma} + \delta_{q \tau}^{\sigma},
\end{equation}
where the radiative shifts are given by
\begin{align}
\label{eq:shift_chain}
\delta_{q\tau}^\sigma=&\;\frac{\eta_\sigma}{4}\frac{\omega_0^2}{\omega_{q\tau}^\sigma}\frac{q^2a^3}{d}\mathcal{B}_{q\tau}^\sigma\Theta(\omega_\mathrm{c}-c|q|)
\left\{ \ln \left( \frac{\omega_\mathrm{c}}{c |q|} \right)
 \right.
 \nonumber\\
 &\left.+ \frac{1}{2} \left[ 1 + \mathrm{sgn} \{ \eta_{\sigma} \} \left( \frac{\omega_{q \tau}^{\sigma}}{cq} \right)^2 \right] 
 \ln \left( \frac{ |(cq)^2 - {\omega_{q \tau}^{\sigma}}^2 | }{\omega_\mathrm{c}^2 - {\omega_{q \tau}^{\sigma}}^2} \right) \right\}.
\end{align}
Here, $\Theta(z)$ is the Heaviside step function and the ultraviolet cutoff $\omega_\mathrm{c}$ is of the order of $c/a$, corresponding to the frequency above which the dipolar approximation used in equation \eqref{eq:HamCoupling} breaks down.
Notably, the frequency shift \eqref{eq:shift_chain} is only logarithmically dependent on the cutoff frequency $\omega_\mathrm{c}$. This is reminiscent of the so-called Bethe logarithm in the equivalent expression for the Lamb shift in atomic physics~\cite{Bethe1947, Milonni1994}. Also immediately apparent from equation \eqref{eq:shift_chain} is the characteristic size dependence $\delta_{q \tau}^{\sigma} \propto a^3$, as is the case of the regular chain~\cite{Downing2018}.

\begin{figure}[tb]
 \includegraphics[width=1.0\columnwidth]{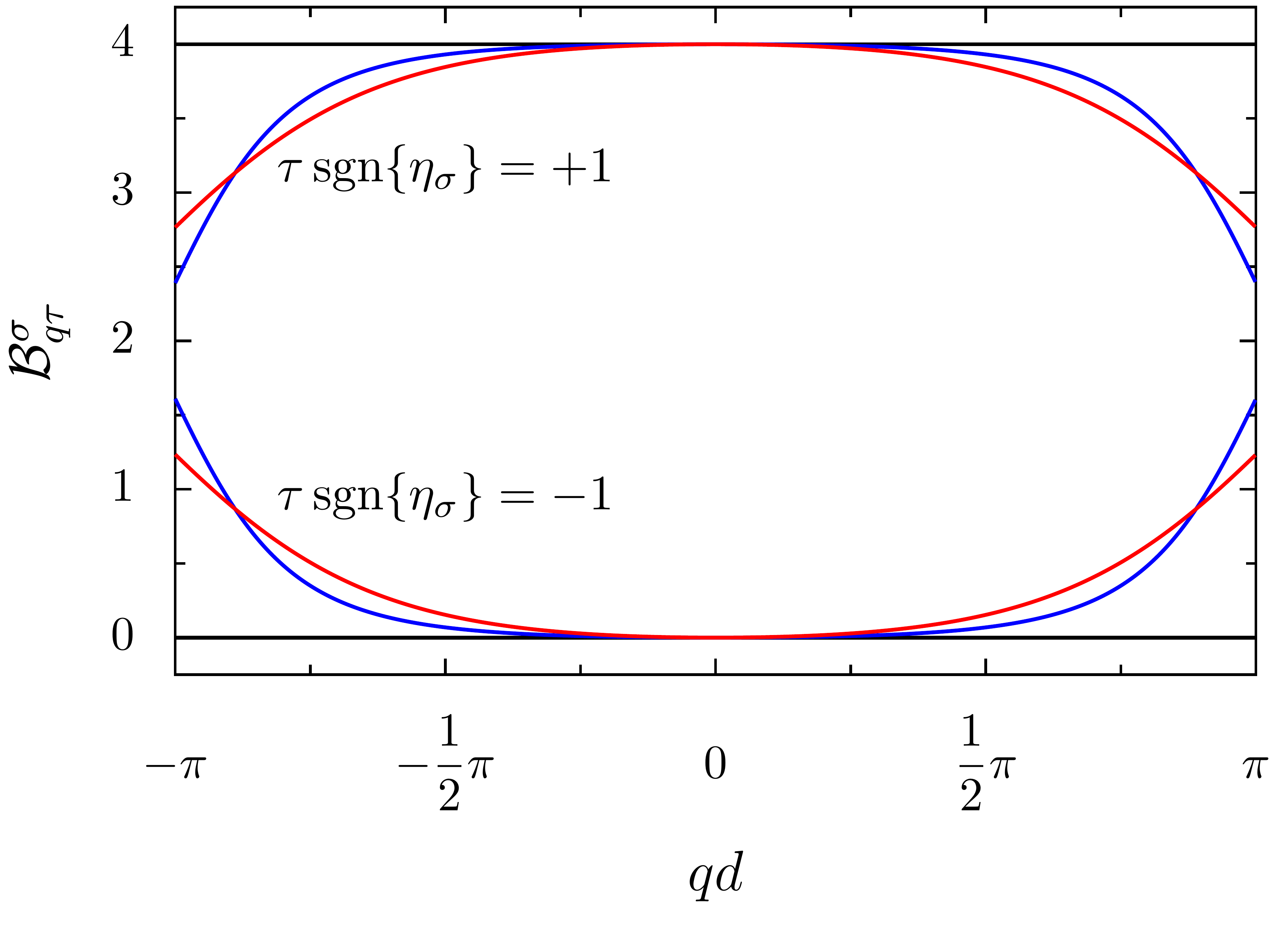}
 \caption{Brightness factor $\mathcal{B}_{q\tau}^\sigma$ from equation \eqref{eq:bright} in the first Brillouin zone for $|\epsilon|=0$ (undimerized case, black lines), $|\epsilon|=1/8$ (blue lines), and $|\epsilon|=1/4$ (red lines). The periodicity of the lattice is $d=8a$.
 }
 \label{fig:brightness}
\end{figure}

\begin{figure*}[tb]
 \includegraphics[width=1.0\linewidth]{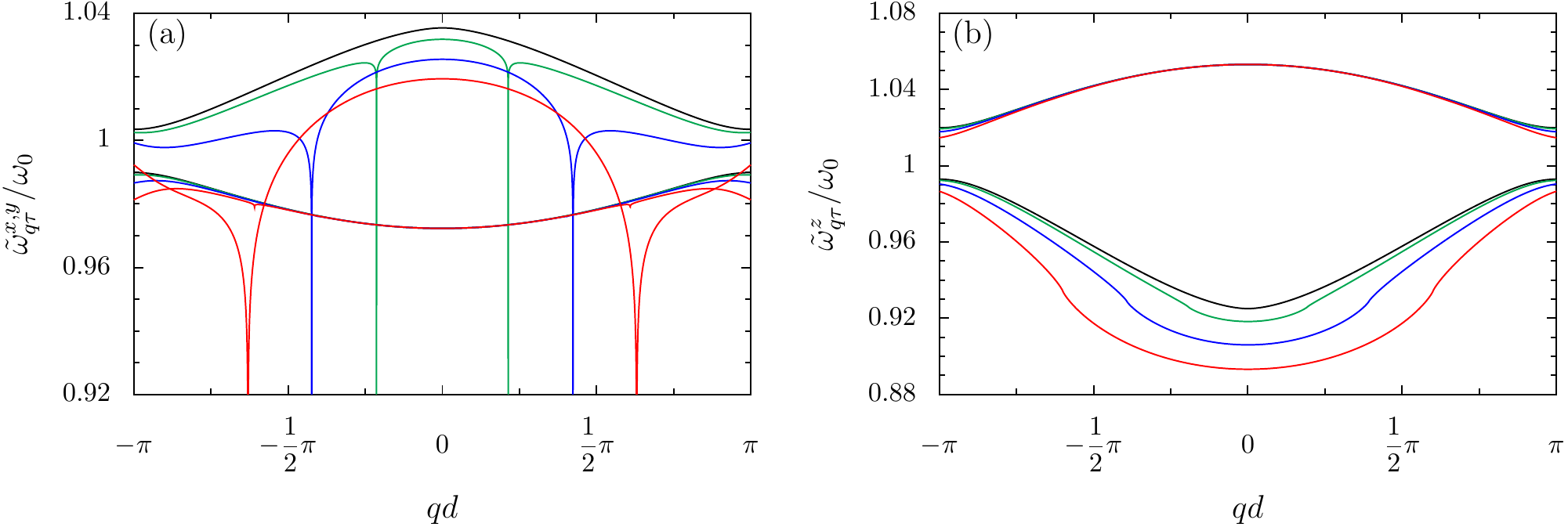}
 \caption{Collective plasmonic bandstructure, in units of the bare LSP resonance frequency $\omega_0$, shown in the first Brillouin zone as a function of the reduced plasmonic wavevector $q d$. Both the (a) transverse and (b) longitudinal polarizations are displayed. Black lines: quasistatic dispersion relation $\omega_{q \tau}^{\sigma}$, neglecting coupling to the photonic environment (cf.\ equation \eqref{eq:plspectrum}). Colored lines: plasmonic dispersion relation $\tilde{\omega}_{q \tau}^{\sigma}$, including coupling to the photonic environment (cf.\ equation \eqref{renorm} with equation \eqref{eq:shift_chain}) for the reduced nanoparticle radii $k_0 a =0.1$ (green lines), $k_0 a = 0.2$ (blue lines), and $k_0 a = 0.3$ (red lines). The parameters used in the figure are: dimerization parameter $|\epsilon|=1/13$, periodicity $d=6.5a$, and ultraviolet cutoff frequency $\omega_\mathrm{c} = c/a$.}
 \label{fig:shift}
\end{figure*}

In equation \eqref{eq:shift_chain}, we have introduced the brightness factor 
\begin{equation}
\label{eq:bright}
\mathcal{B}_{q\tau}^\sigma=\left|
1+\tau\,\mathrm{e}^{\mathrm{i}(qd_1+\phi_q^\sigma)}
\right|^2,
\end{equation}
where the phase $\phi_q^\sigma$ is defined in equation \eqref{eq:phi}. 
This quantity measures the strength of the light-matter coupling, and arises from the first term on the right-hand side of the coupling Hamiltonian 
\eqref{eq:HamCoupling}. Importantly, $\mathcal{B}_{q\tau}^\sigma$ depends on the polarization $\sigma$ and the band index $\tau$ only through the product $\tau\,\mathrm{sgn}\{\eta_\sigma\}$. This is exemplified in figure \ref{fig:brightness}, where we show the brightness factor in the first Brillouin zone for several values of the dimerization parameter $|\epsilon|$. As can be seen from the figure, there are drastically different behaviors of $\mathcal{B}_{q\tau}^\sigma$ for $\tau\,\mathrm{sgn}\{\eta_\sigma\}=+1$ ($-1$) which corresponds to bright (dark) plasmonic bands.
Therefore, for the transverse polarization ($\sigma = x, y$) the high-energy band ($\tau = +$) is bright (i.e.\ it significantly couples to light) and corresponds to an orientation where the dipole moments on each nanoparticle in a given dimer are in-phase ($\uparrow \uparrow$). Whereas, the low-energy band ($\tau = -$) refers to a situation where the dipole moments on each nanoparticle in a dimer are out-of-phase ($\uparrow \downarrow$), or a dark band. 
For the longitudinal polarization ($\sigma = z$), the aformentioned situation is inverted: $\tau=+$ ($\tau=-$) corresponds to a dark (bright) 
band.\footnote{Whether a band is bright or dark can in fact be easily envisioned, at least for the $q=0$ mode, from the very nature of the dipole-dipole interaction, which energetically favors an antiparallel (parallel) alignment of the dipole moments in the transverse (longitudinal) configuration.}
Notice that for plasmonic modes with wavevectors close to the Brillouin zone edge, this classification is less pronounced in the dimerized case ($|\epsilon|\neq0$).
 
The strikingly different polarization-dependent behavior of $\mathcal{B}_{q\tau}^\sigma$ shown in figure \ref{fig:brightness} has important implications for the frequency shifts \eqref{eq:shift_chain}, as well as for the radiative decay rates of the system (see appendix \ref{app:decay_rates}).
In figure \ref{fig:shift}, we plot the renormalized plasmonic bandstructure \eqref{renorm} for both the transverse (panel (a)) and longitudinal polarizations (panel (b)) as colored lines, for nanoparticle radii $k_0 a =0.1$ (green lines), $k_0 a = 0.2$ (blue lines), $k_0 a =0.3$ (red lines) with $k_0=\omega_0/c$, and fixed periodicity $d=6.5a$. 
In the figure, the dimerization parameter is $|\epsilon|=1/13$ (cf.\ equation \eqref{eq:epsilon}) and the ultraviolet cutoff frequency is chosen to be $\omega_\mathrm{c} = c/a$.
With increasing nanoparticle size, the radiative corrections \eqref{eq:shift_chain} significantly modify the quasistatic result in figure \ref{fig:shift} (see black lines) for the bright bands and are negligible for the dark bands. Thus, for the transverse polarization it is the upper-in-energy ($\tau=+$) band which experiences a sizable redshift, while for the longitudinal polarization it is the lower band ($\tau=-$). 

Most noticeable from figure \ref{fig:shift} are the presence of cusps in the dispersion relation of the upper band of the transverse modes (panel (a)). In the large-chain limit ($\mathcal{N}\gg1$), such cusps corresponds to logarithmic singularities, see the last term in equation \eqref{eq:shift_chain}. These singularities occurs at the intersection between the plasmonic dispersion and the light cone ($\omega_{q +}^{x, y} = c |q|$), i.e.\ at $q \simeq \pm k_0$ to zero order in $\Omega/\omega_0\ll1$.\footnote{Formally, there are also logarithmic singularities in the lower transverse band at $\omega_{q -}^{x, y} = c |q|$. However, this is a dark band so that these singularities are heavily suppressed by the near-zero behavior of the  brightness factor $\mathcal{B}_{q\tau}^\sigma$ shown in figure \ref{fig:brightness}.}  These striking features also manifest in the associated radiative damping, where they appear as discontinuities in the decay rate at the same points in wavevector space (see appendix~\ref{app:decay_rates}). Remarkably, the influence of the cusp singularities on the bandstructure leads to a band inversion for larger particles (red lines in figure \ref{fig:shift}(a)), since the $\tau = +$ band dips below the $\tau = -$ band in some regions of the first Brillouin zone. 
However, this band crossing phenomenon should be treated with some caution, since it arises at the limit of applicability of our dipolar approximation (cf.\ equation \eqref{eq:HamCoupling}).

The behavior observed in figure \ref{fig:shift} and encapsulated in equations \eqref{renorm} and \eqref{eq:shift_chain} has been recently reported by means of sophisticated numerical calculations based on solutions of the fully-retarded Maxwell equations \cite{Pocock2018}. Our elementary open quantum system approach provides a simple and transparent analytical understanding of the plasmonic bandstructure, including the intriguing effects of retardation. 

Radiative shifts and radiative decay rates are intrinsically linked by the fluctuation-dissipation theorem \cite{Kubo1966}. 
The reporting of the exact result of the quasistatic bandstructure~\eqref{eq:plspectrum} means that the accuracy of the radiation damping decay rate, previously derived within the nearest-neighbor approximation in reference \cite{DowningBiPartite2017}, can be improved upon. In appendix \ref{app:decay_rates}, we provide the exact expression for the radiation damping decay rate including the effects of long-range interactions. 

Since the radiation damping decay rate is vanishing outside the light cone, that is for wavevectors $|q| \gtrsim k_0$, the edge states discussed in section \ref{Sec:topological} are dark (due to being pinned at $|q| =  \pi /d$). It then follows that these edge states weakly interact with light and so should not be destroyed by retardation effects. Indeed, while the inclusion of retardation acts to further break the band reversal (particle-hole) and chiral symmetries of the collective plasmons, these broken symmetries were already shown in section \ref{Sec:topological} to be inconsequential for the existence of topologically nontrivial states. Then, as long as there is a well-defined bandgap at the Brillouin zone edge, plasmonic edges states should be detectable. This physical argument is supported by the latest numerical calculations on finite plasmonic chains with retardation~\cite{Pocock2018}, where it is also shown that plasmonic edge states are robust to moderate disorder.

Finally, we emphasize that the results presented in this section arise from a perturbative calculation. 
Consequently, the singularity in the transverse collective plasmon dispersion, 
leading to a large deviation from the natural LSP frequency $\omega_0$, should be treated with caution.
An analysis of the strong coupling regime, where the quasiparticles are plasmon-polaritons \cite{Torma2015}, is beyond the scope of this work.

%==================================================
%==================================================
%==================================================
\section{Conclusions}
\label{sec:conc}

We have presented a simple Hamiltonian model of collective plasmons in a dimerized chain of spherical metallic nanoparticles. Taking into account long-range quasistatic interactions, we have derived an exact expression for the plasmonic bandstructure. 
We have shown that the Bloch Hamiltonian obeys time-reversal and inversion symmetries, such that the Zak phase is quantized, despite the broken 
particle-hole and chiral symmetries of the system. 
We have calculated the Zak phase explicitly, 
finding a topologically nontrivial regime of dimerization, which proves the existence of plasmonic edge states which are 
robust against long-range interactions, bosonic statistics, and nonresonant couplings.

We have also taken into account retardation effects and calculated analytical expressions for the resulting radiative frequency shifts. 
While the dark bands are essentially unaffected by retardation, the bright bands present a size-dependent frequency shift. The latter is most pronounced for the transverse-polarized modes in the vicinity of the intersection between the quasistatic spectrum and the light cone.  Otherwise, 
the fully-retarded plasmonic bandstructure presents only a quantitative difference with respect to the quasistatic result for small nanoparticles. 
Therefore, the topologically nontrivial edge states are essentially unaffected by retardation.

%==================================================
%==================================================
%==================================================
%==================================================
\section*{Acknowledgments}

We thank Fran\c{c}ois Gautier and Dietmar Weinmann for useful discussions. We acknowledge financial support from the French Agence Nationale de la Recherche (Project ANR-14-CE26-0005 Q-MetaMat) and from the 
Spanish Ministerio de Econom\'ia y Competitividad (Project FIS2015-64951-R CLAQUE).

%==================================================
%==================================================
%==================================================
%==================================================
\section*{Author contribution statement}
Both authors contributed equally to this paper.

\appendix
%\setcounter{equation}{0}%reset counter
%\renewcommand{\theequation}{A\arabic{equation}}

%==================================================
%==================================================
%==================================================
%==================================================
\section{Lattice sums $f_q^\sigma$ and $g_q^\sigma$}
\label{app:lattice_sums}
In this appendix we discuss the behavior of the lattice sums \eqref{eq:fsum} and \eqref{eq:gsum}, which are both $2\pi/d$-periodic as a function of the wavevector $q$. 

\begin{figure}[tb]
 \includegraphics[width=1.0\linewidth]{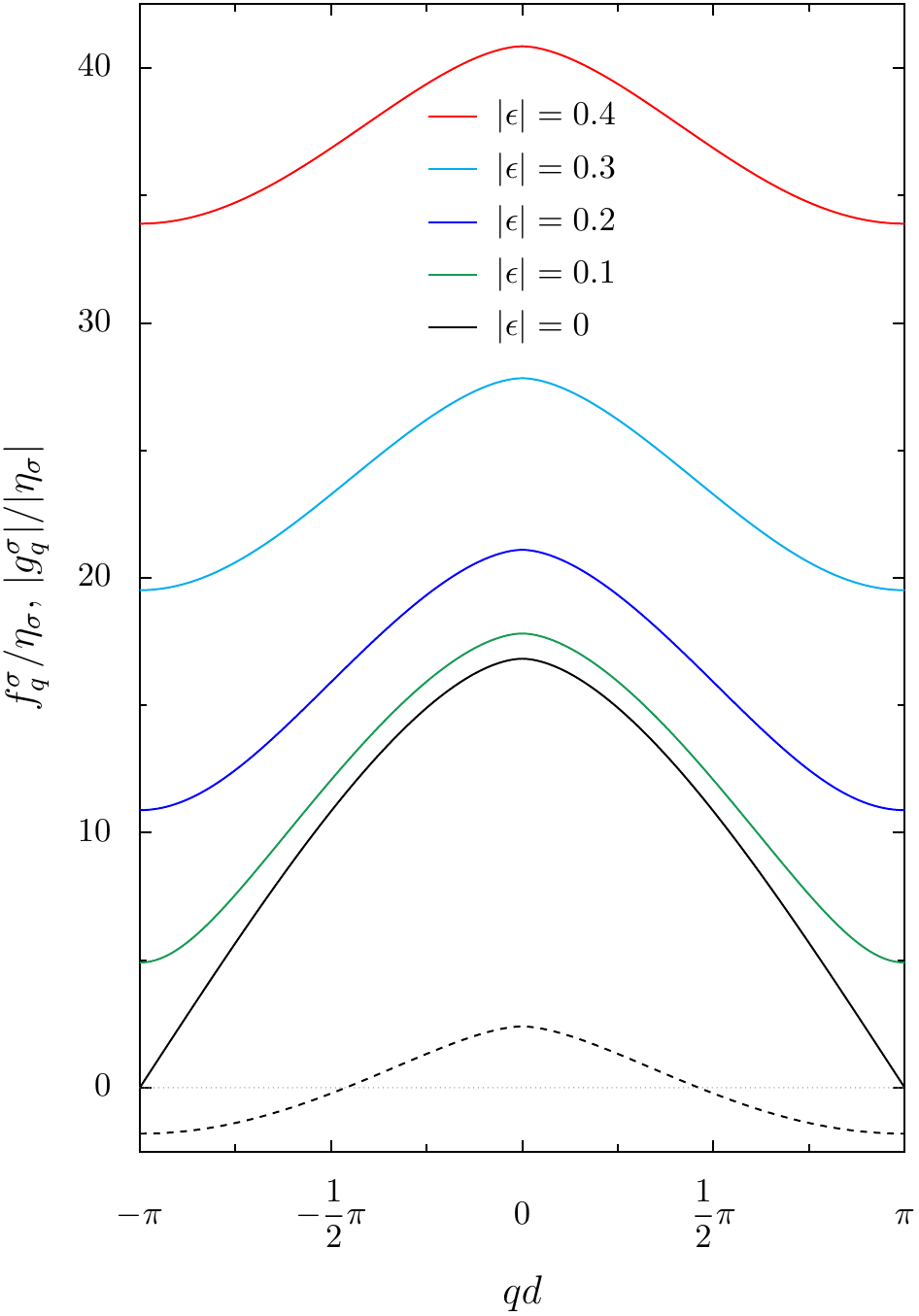}
 \caption{Dashed line: lattice sum $f_q^\sigma$ scaled by $\eta_\sigma$ as a function of $qd$ (cf.\ equation \eqref{eq:fsum}). Solid lines: absolute value of the lattice sum $g_q^\sigma$ (cf.\ equation \eqref{eq:gsum}) in units of $|\eta_\sigma|$ as a function of $qd$ and for increasing absolute values of the dimerization parameter $\epsilon$, defined in equation \eqref{eq:epsilon}.}
 \label{fig:lattice_sums}
\end{figure}

The even intrasublattice function $f_q^\sigma$ is plotted as a dashed line in figure \ref{fig:lattice_sums}. 
It is bounded by its values at the center and edge of the first Brillouin zone, $f_{0}^{\sigma} = 2  \zeta (3) \eta_{\sigma}$ and 
$f_{\pi /d}^{\sigma} = -3 \zeta (3) \eta_{\sigma}/2$,
respectively, 
where Ap\'{e}ry's constant $\zeta (3) \simeq 1.20$ is defined through the Riemann zeta function $\zeta (s) = \sum_{n=1}^{\infty} n^{-s}$. 
In the vicinity of the center of the first Brillouin zone ($|q|d\ll1$), we have $f_q^\sigma/\eta_\sigma\simeq2\zeta(3)-(3-2\ln|q|)q^2/2$, while close to the Brillouin zone edge, $f_q^\sigma/\eta_\sigma\simeq-3\zeta(3)/2+\ln{(2)} (|q|d-\pi)^2$.

The absolute value of the intersublattice function $g_q^\sigma$ which enters the quasistatic spectrum \eqref{eq:plspectrum} is also even and is plotted as solid lines in figure \ref{fig:lattice_sums} for increasing absolute values of the dimerization parameter $\epsilon$ defined in equation \eqref{eq:epsilon}. Note that $|g_q^\sigma|$ is an even function of the parameter $\epsilon\in[-1, +1]$.
The intersublattice function~\eqref{eq:gsum} has the boundary values
\begin{equation}
\label{eq:gbounds1}
  g_{0}^{\sigma} = \eta_{\sigma} \left[ \zeta (3, d_1/d) + \zeta (3, d_2/d) \right]
\end{equation}
and
\begin{align}
\label{eq:gbounds2}
 g_{\pi /d}^{\sigma} =&\; \frac{\eta_{\sigma}}{8} \left\{ \zeta (3, {d_1}/{2 d}) + \zeta (3, {[d + d_2]}/{2 d})  \right.\nonumber\\
 &- \left. \zeta (3, {d_2}/{2 d}) - \zeta (3, {[d + d_1]}/ {2 d}) \right\},
  \end{align}
where the Hurwitz zeta function is defined as
$\zeta (s, z) = \sum_{n=0}^{\infty}(n+z)^{-s}$.

%==================================================
%==================================================
%==================================================
%==================================================
\section{Mapping of the quasistatic plasmonic bandstructure at the Brillouin zone edge to a quasirelativistic spectrum}
\label{app:mapping}
Here we provide details of the expansion of the quasistatic spectrum \eqref{eq:plspectrum} close to the edge of the first Brillouin zone, showing that it maps to the quasirelativistic spectrum \eqref{eq:edgeBZspectrum}. 

To leading order in $kd\ll1$, with $qd=kd+\pi$, we find after a lengthy but straightforward Taylor expansion, that equation \eqref{eq:plspectrum} reduces to equation \eqref{eq:edgeBZspectrum}, with 
\begin{equation}
\label{eq:coeffA}
 A^{\sigma} = | g_{\pi /d}^{\sigma} |^2,
\end{equation}
where $g_{\pi /d}^{\sigma}$ is given in equation \eqref{eq:gbounds2}, and
\begin{align}
\label{eq:coeffB}
 B^{\sigma} =&\; |\eta_{\sigma}|^2 \big\{ B_1(d_1/d) B_1(d_2/d)  \nonumber\\
&+B_1(d_1/d) B_2(d_1/d, d_2/d) \nonumber\\
&+ B_1(d_2/d) B_2(d_2/d, d_1/d)  \nonumber\\
 &+  \left[ B_3(d_1/d)+B_3(d_2/d) \right]^2 \big\}.
\end{align}
Here we have introduced the auxiliary functions
\begin{subequations}
\begin{equation}
 B_1(x) = \frac{1}{8} \left\{ \zeta(3, x/2) - \zeta(3, [1+x]/2) \right\},
\end{equation}
\begin{align}
 B_2(x, y) =&\; \frac{1}{(1+x)^3} - \frac{4}{(2+x)^3} +  \frac{9}{(3+x)^3} \nonumber\\
 &- \frac{3}{(1+y)^3} + \frac{8}{(2+y)^3} -  \frac{15}{(3+y)^3},
\end{align}
and
\begin{equation}
 B_3(x) = \frac{1}{(1+x)^3} - \frac{2}{(2+x)^3} +  \frac{3}{(3+x)^3},
\end{equation}
\end{subequations}
where the Hurwitz zeta function $\zeta (s, z)$ is defined in appendix~\ref{app:lattice_sums}. 
This completes the analytical description of the quasirelativistic plasmonic dispersion at the Brillouin zone edge~\eqref{eq:edgeBZspectrum}.

%==================================================
%==================================================
%==================================================
%==================================================
\section{Photonic-induced frequency shifts of the collective plasmons}
\label{app:shifts}
In this appendix, we provide details of the calculation of the renormalized plasmonic dispersion \eqref{renorm} and the radiative frequency shifts \eqref{eq:shift_chain} induced by the photonic bath in which the nanoparticles are embedded. 

Following reference \cite{Downing2018}, we treat the plasmon-photon coupling Hamiltonian \eqref{eq:HamCoupling} up to second order in perturbation theory. For a given mode of polarization $\sigma$, band index $\tau$, and wavevector $q$, the plasmonic energy levels are
\begin{equation}
\label{eq:E00}
E_{n_{q \tau}^\sigma} = E_{n_{q \tau}^\sigma}^{(0)} + E_{n_{q \tau}^\sigma}^{(1)} + E_{n_{q \tau}^\sigma}^{(2)},
\end{equation}
where $n_{q \tau}^\sigma$ is a non-negative integer. 
Here, the zeroth-order contribution $E_{n_{q \tau}^\sigma}^{(0)}=n_{q \tau}^{\sigma} \hbar \omega_{q \tau}^{\sigma}$ is simply the unperturbed quantum harmonic oscillator levels corresponding to the Hamiltonian \eqref{eq:plchain}, where the frequency $\omega_{q \tau}^{\sigma}$ is given in equation \eqref{eq:plspectrum}.

The first-order contribution $E_{n_{q \tau}^\sigma}^{(1)}$ arises from the second, ``diamagnetic'' term on the right-hand side of equation \eqref{eq:HamCoupling}. It corresponds to the process whereby a virtual photon, with wavevector $\mathbf{k}$ and transverse polarization $\hat\lambda_\mathbf{k}$, is emitted and absorbed. Notably, this self-energy term does not involve any plasmonic operators and so the plasmonic eigenstate $|n_{q \tau}^\sigma\rangle$ remains unchanged by the process. Explicitly, one finds 
\begin{equation}
\label{eq:E1}
E_{n_{q \tau}^\sigma}^{(1)} = 4 \pi \mathcal{N} \hbar\omega_0^2\frac{a^3}{\mathcal{V}}\sum_\mathbf{k}\frac{1}{\omega_\mathbf{k}},
\end{equation}
which corresponds to a global energy shift, independent of the quantum number $n_{q \tau}^\sigma$. Since only interlevel energy differences are observable through plasmonic transitions, this seemingly divergent contribution does not lead to a renormalization of the bare collective mode resonance frequency $\omega_{q\tau}^\sigma$ and may be discarded.

The second-order contribution $E_{n_{q \tau}^\sigma}^{(2)}$ in equation \eqref{eq:E00} arises from the first, ``paramagnetic'' term on the right-hand side of the coupling Hamiltonian~\eqref{eq:HamCoupling}. It corresponds to the emission and reabsorption of virtual photons via the intermediate plasmonic states $|n_{q \tau}^\sigma \pm 1\rangle$. One obtains
\begin{align}
\label{eq:secondorder}
 E_{n_{q \tau}^\sigma}^{(2)} =&\; \frac{\pi}{2} \hbar \omega_0^2\omega_{q\tau}^\sigma \frac{a^3}{\mathcal{V}} \sum_{\mathbf{k}, \hat{\lambda}_{\mathbf{k}}} \frac{|\hat{\sigma} \cdot \hat{\lambda}_{\mathbf{k}} |^2}{\omega_{\mathbf{k}}} \nonumber\\
 &\times  \left\{ \left\vert 1 + \tau\, \mathrm{e}^{\mathrm{i} (k_z d_1+\phi_q^\sigma)} \right\vert^2
  \frac{n_{q \tau}^\sigma  |F_{\mathbf{k}, q}^{-}|^2}{\omega_{q \tau}^{\sigma} - \omega_{\mathbf{k}}} \right.\nonumber\\
 &- \left.\left\vert 1 + \tau\, \mathrm{e}^{\mathrm{i} (k_z d_1-\phi_q^\sigma)} \right\vert^2 \frac{(n_{q \tau}^\sigma + 1) |F_{\mathbf{k}, q}^{+}|^2 }{\omega_{q \tau}^{\sigma} + \omega_{\mathbf{k}}} \right\},
\end{align}
where the summation over $\mathbf{k}$ excludes the singular term for which $\omega_{\mathbf{k}} = \omega_{q \tau}^{\sigma}$. In the expression above, the array factor reads as 
\begin{equation}
\label{eq:array}
 F_{\mathbf{k}, q}^{\pm} =\frac{\mathrm{e}^{\mp\mathrm{i}k_zd}}{\sqrt{ \mathcal{N}}}
 \sum_{n=1}^\mathcal{N} \mathrm{e}^{\mathrm{i} n (q\pm k_z )d},
\end{equation}
with $k_z=\mathbf{k}\cdot\hat z$, and the phase $\phi_q^\sigma$ is defined in equation \eqref{eq:phi}.
In the continuum limit, where $\sum_{\mathbf{k}} \to \mathcal{P} \int \mathrm{d}^3 \mathbf{k} \; \mathcal{V}/{(2 \pi)^3}$ (here, $\mathcal{P}$ denotes the Cauchy principal value), the second-order correction \eqref{eq:secondorder} is divergent. This divergence can be regularized by introducing a physically-motivated ultraviolet cutoff $k_\mathrm{c}\sim 1/a$, corresponding to the wavelength below which the dipolar approximation used in equation \eqref{eq:HamCoupling} breaks down.

The renormalized frequency difference between successive plasmonic energy levels $\tilde{\omega}_{q \tau}^{\sigma} = (E_{n_{q \tau}^{\sigma}+1} - E_{n_{q \tau}^{\sigma}})/\hbar$ is then independent of the quantum number $n_{q \tau}^{\sigma}$, up to second order in perturbation theory. 
Hence, the radiative frequency shift (see equation \eqref{renorm}) is given by
\begin{align}
\label{eqS:lambsingle}
 \delta_{q \tau}^{\sigma} =&\; \frac{\pi}{2} \omega_0^2\omega_{q\tau}^\sigma \frac{a^3}{\mathcal{V}}  \sum_{\mathbf{k}, \hat{\lambda}_{\mathbf{k}}} \frac{|\hat{\sigma} \cdot \hat{\lambda}_{\mathbf{k}} |^2}{\omega_{\mathbf{k}}} \nonumber\\
 &\times  \left\{ \left\vert 1 + \tau\, \mathrm{e}^{\mathrm{i} (k_z d_1+\phi_q^\sigma)}\right\vert^2 \frac{ |F_{\mathbf{k}, q}^{-}|^2}{\omega_{q \tau}^{\sigma} - \omega_{\mathbf{k}}} \right.\nonumber\\
 &- \left. \left\vert 1 + \tau\,  \mathrm{e}^{\mathrm{i} (k_z d_1-\phi_q^\sigma)} \right\vert^2 \frac{|F_{\mathbf{k}, q}^{+}|^2}{\omega_{q \tau}^{\sigma} + \omega_{\mathbf{k}}} \right\}.
\end{align}
The  identity
\begin{equation}
\label{eq:sum_polarization}
\sum_{\hat{\lambda}_{\mathbf{k}}}|\hat{\sigma} \cdot \hat{\lambda}_{\mathbf{k}} |^2 = 1 - ( \hat{\sigma}\cdot {\hat{k}} )^2
\end{equation}
 allows one to perform the summation over photon polarization in equation \eqref{eqS:lambsingle}. Transforming the wavevector summation into a principal-value integral, we obtain in spherical coordinates $(k, \theta, \varphi)$
\begin{align}
\label{eq:delta_shift2}
\delta_{q \tau}^{\sigma} =&\; \frac{1}{16\pi^2c}\omega_0^2\omega_{q\tau}^\sigma a^3\ \mathcal{P}\int_0^{k_\mathrm{c}}  \mathrm{d}k\, k \nonumber\\
 &\times \int_0^\pi \mathrm{d}\theta\,\sin{\theta}\left( \left\vert 1 + 
 \tau\,\mathrm{e}^{\mathrm{i} (k\cos{\theta} d_1+\phi_q^\sigma)} \right\vert^2 \frac{|F_{\mathbf{k}, q}^{-}|^2 }{\omega_{q \tau}^{\sigma} - \omega_{\mathbf{k}}} \right.\nonumber\\
 &- \left. \left\vert 1 + \tau\,\mathrm{e}^{\mathrm{i} (k\cos{\theta} d_1-\phi_q^\sigma)} \right\vert^2 \frac{ |F_{\mathbf{k}, q}^{+}|^2}{\omega_{q \tau}^{\sigma} + \omega_{\mathbf{k}}} \right) \nonumber\\
 &\times \int_0^{2\pi} \mathrm{d}\varphi[1-(\hat k\cdot\hat\sigma)^2] ,
\end{align}
where the cutoff $k_\mathrm{c} > \omega_{q\tau}^{\sigma} /c$. The integral over the azimuthal angle $\varphi$ is readily evaluated with the identity
\begin{equation}
\label{eq:int_phi}
\int_0^{2\pi}  \mathrm{d}\varphi[1-(\hat k\cdot\hat\sigma)^2] =
\pi|\eta_\sigma|\left(1+\mathrm{sgn}\{\eta_\sigma\}\cos^2{\theta}\right).
\end{equation}
Since we are working in the long-chain regime ($\mathcal{N}\gg1$), the subsequent integral over the polar angle $\theta$ 
is easily obtained using the result 
\begin{equation}
\label{eq:F_long}
|F_{\mathbf{k}, q}^{\pm}|^2 \simeq  2 \pi \delta \left( [q \pm k\cos\theta] d\right),
\end{equation}
where $\delta (z)$ is the Dirac delta function. The final integral over the radial coordinate $k$ then yields the desired result~\eqref{eq:shift_chain}.

%==================================================
%==================================================
%==================================================
%==================================================
\section{Radiation damping decay rates}
\label{app:decay_rates}

\begin{figure*}[tb]
 \includegraphics[width=1.0\linewidth]{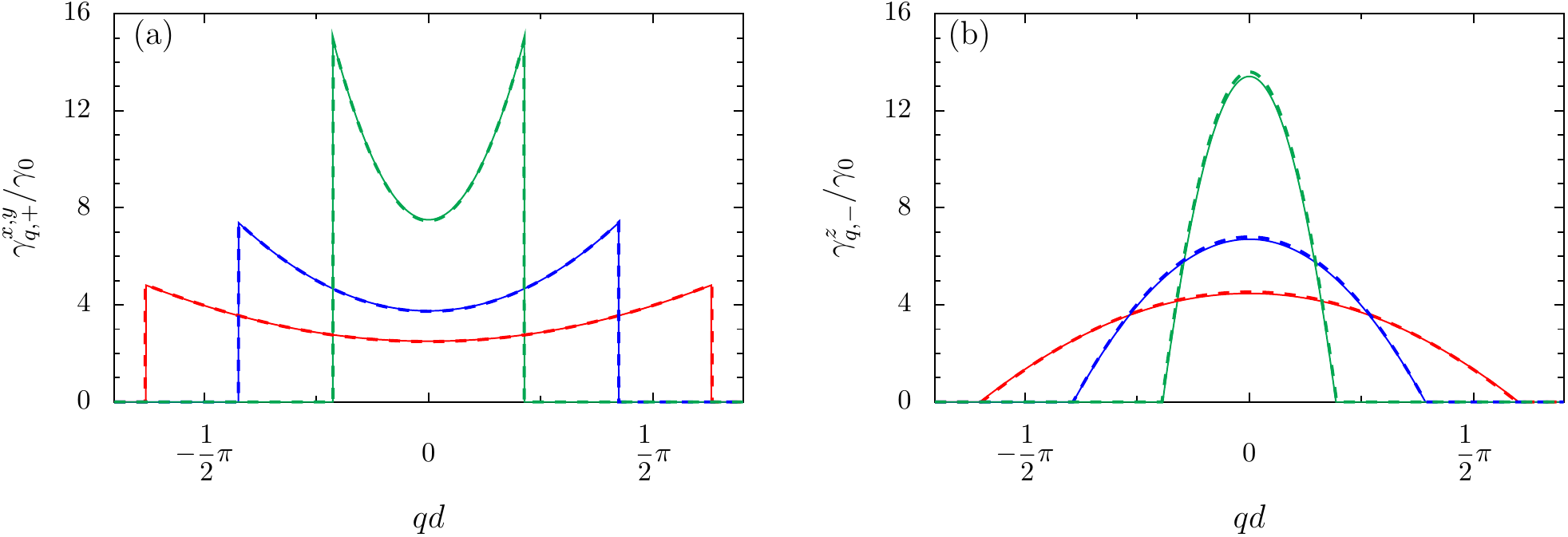}
 \caption{Radiation damping decay rate of the bright bands, in units of the single particle radiative rate $\gamma_0$, as a function of $q d$ and for (a) the transverse and (b) longitudinal polarizations. Solid lines: radiation damping decay rate $\gamma_{q \tau}^{\sigma}$ including the long-ranged dipole-dipole interaction, see equation \eqref{eq:damp_new}. Dashed lines: radiation damping decay rate in the nearest-neighbor approximation, see equation (33) in reference \cite{DowningBiPartite2017}. In the figure, we show results for the reduced nanoparticle sizes $k_0 a =0.1$ (green lines), $k_0 a = 0.2$ (blue lines), and $k_0 a = 0.3$ (red lines). The periodicity and the dimerization of the chain are $d=6.5a$ and $|\epsilon|=1/13$, respectively.}
 \label{fig:damping}
\end{figure*}

Here we present a detailed calculation of the radiation damping decay rates of the collective plasmons, including the long-ranged quasistatic dipole-dipole interaction, thereby extending the results of reference \cite{DowningBiPartite2017} which was restricted to nearest-neighbor interactions only. 

The radiative damping of a $\sigma$-polarized collective mode with wavevector $q$ in the band $\tau$ arises from the decay of such a state, resulting in the spontaneous emission of photons with momentum $\mathbf{k}$ and transverse polarization $\hat\lambda_\mathbf{k}$ in the far field. 
The associated radiation damping decay rates $\gamma_{q\tau}^\sigma$ of the system follow directly from a Fermi golden rule calculation with the coupling Hamiltonian \eqref{eq:HamCoupling}, yielding
\begin{align}
\label{eq:gamma_FGR}
\gamma_{q\tau}^\sigma=\;&
\pi^2\omega_0^2\omega_{q\tau}^\sigma\frac{a^3}{\mathcal{V}} \sum_{\mathbf{k}, \hat{\lambda}_{\mathbf{k}}} \frac{|\hat{\sigma} \cdot \hat{\lambda}_{\mathbf{k}} |^2}{\omega_{\mathbf{k}}}
\left\vert 1 + \tau\, \mathrm{e}^{\mathrm{i} (k_z d_1+\phi_q^\sigma)}  \right\vert^2
\nonumber\\
&\times
|F_{\mathbf{k},q}^-|^2\delta(\omega_{q\tau}^\sigma-\omega_\mathbf{k}), 
\end{align}
where the array factor $F_{\mathbf{k},q}^-$ is given in equation \eqref{eq:array} and the phase $\phi_{q}^\sigma$ is defined in equation \eqref{eq:phi}.
In the continuum limit, we replace the summation over photonic wavevectors by an integral, and using equation \eqref{eq:sum_polarization}, we obtain
\begin{align}
\gamma_{q \tau}^{\sigma} =&\; \frac{1}{8\pi c}\omega_0^2\omega_{q\tau}^\sigma a^3\int_0^{\infty}  \mathrm{d}k\, k\ \delta(\omega_{q\tau}^\sigma-ck)
\int_0^\pi \mathrm{d}\theta\,\sin{\theta}
 \nonumber\\
 &\times\left\vert 1 + \tau\,\mathrm{e}^{\mathrm{i} (k\cos{\theta} d_1+\phi_q^\sigma)} \right\vert^2
 |F_{\mathbf{k}, q}^{-}|^2
 \nonumber\\
&\times\int_0^{2\pi} \mathrm{d}\varphi[1-(\hat k\cdot\hat\sigma)^2].
\end{align}
It should be noted that the $k$-integral in the above equation does not require the introduction of the ultraviolet cutoff $k_\mathrm{c}$, contrary to the case of the radiative frequency shifts (see equation \eqref{eq:delta_shift2}). This is because the Dirac delta function in equation \eqref{eq:gamma_FGR} imposes conservation of energy, which is compatible with the dipolar approximation used in equation \eqref{eq:HamCoupling}. 
With the help of equations \eqref{eq:int_phi} and \eqref{eq:F_long}, we then obtain in the long-chain limit 
($\mathcal{N}\gg1$) the final result
\begin{align}
\label{eq:damp_new}
\gamma_{q\tau}^\sigma=\;&\frac{\pi\eta_\sigma}{4}\frac{\omega_0^2}{\omega_{q\tau}^\sigma}\frac{q^2a^3}{d}
\mathcal{B}_{q\tau}^\sigma
\Theta(\omega_{q\tau}^\sigma-c|q|)
\nonumber\\
&\times\left[1+\mathrm{sgn}\{\eta_\sigma\}\left(\frac{\omega_{q\tau}^\sigma}{cq}\right)^2\right], 
\end{align}
where the brightness factor $\mathcal{B}_{q\tau}^\sigma$ is defined in equation \eqref{eq:bright}. 
The expression above presents an $a^3$-dependence, as is the case for the radiative damping of a single nanoparticle $\gamma_0=2\omega_0^4a^3/3c^3$. 
Immediately apparent in the above expression is the influence of the light cone at $|q| \simeq k_0$ (through the Heaviside step function $\Theta (z)$) as was noticed for the associated radiative shifts~\eqref{eq:shift_chain}. Clearly, the decay rate is vanishing outside the light cone ($|q| \gtrsim k_0$). 

In figure \ref{fig:damping}, we plot the radiation damping decay rate \eqref{eq:damp_new} for the bright bands (solid lines) and the same quantity in the nearest-neighbor approximation given in equation (33) of reference \cite{DowningBiPartite2017} (dashed lines).
Due to the brightness factor $\mathcal{B}_{q\tau}^\sigma$ in equation \eqref{eq:damp_new}, the results for the dark bands are negligible in comparison to the bright bands (which display a superradiant behavior), and so we do not plot them. In the figure, we show results for increasing nanoparticle sizes $k_0 a =0.1$ (green lines), $k_0 a = 0.2$ (blue lines), $k_0 a = 0.3$ (red lines), and fixed periodicity $d=6.5a$. The dimerization of the chain is taken to be $|\epsilon|=1/13$. 
Clearly, for the transverse polarization (panel (a)) the two expressions are in excellent agreement, with the dashed lines being barely noticeable. The longitudinal polarization (panel (b)) displays a very good agreement between the exact and approximate expressions, with small deviations only visible at the center of the first Brillouin zone. 
We can therefore conclude that the long-ranged nature of the quasistatic dipole-dipole interaction has almost no quantitative effects on the radiative damping rates of the collective plasmons.

%===========================================================================
%===========================================================================
%===========================================================================
%===========================================================================

\end{document}